**Energy resolution of terahertz single-photon-sensitive bolometric detectors**

D. F. Santavicca,[1] B. Reulet,[2] B. S. Karasik,[3] S. V. Pereverzev,[3] D. Olaya,[4] M. E. Gershenson,[4] L. Frunzio,[1] and D. E. Prober[1]

[1]*Department of Applied Physics, Yale University, New Haven, CT 06520*
[2]*Laboratoire de Physique des Solides, Universite Paris-Sud, 91405 Orsay, France*
[3]*Jet Propulsion Laboratory, California Institute of Technology, Pasadena, CA 91109*
[4]*Department of Physics, Rutgers University, Piscataway, NJ 08854*

We report measurements of the energy resolution of ultra-sensitive superconducting bolometric detectors. The device is a superconducting titanium nanobridge with niobium contacts. A fast microwave pulse is used to simulate a single higher-frequency photon, where the absorbed energy of the pulse is equal to the photon energy. This technique allows precise calibration of the input coupling and avoids problems with unwanted background photons. Present devices have an intrinsic full-width at half-maximum energy resolution of approximately 23 terahertz, near the predicted value due to intrinsic thermal fluctuation noise.

Terahertz (THz) detectors have seen rapid development during the past decade. However, an energy-resolving THz single-photon detector – i.e., a THz calorimeter – has remained elusive. Previous work on semiconductor quantum dot detectors has demonstrated THz single-photon detection, but with a complex device geometry, low quantum efficiency (~1%), and without photon energy resolution.[1,2] The superconducting bolometric detector has the potential to achieve energy-resolved THz single-photon detection with high quantum efficiency in a device with a relatively simple geometry.[3,4]

For a hot electron bolometric calorimeter, with a measurement bandwidth equal to the intrinsic device response bandwidth, the energy resolution is limited by thermodynamic fluctuations, and scales as

$$\delta E_{intrinsic} \sim \sqrt{k_B T^2 C_e} \qquad (1)$$

where $C_e$ is the electronic heat capacity, proportional to the active device volume and the operating temperature T.[5,6] Thus, for sensitive detection, operation is at low temperature and all dimensions of the device are much smaller than a wavelength. Efficient photon coupling can be achieved by integrating the device in a planar THz antenna.[7] An array of such detectors is essential for proposed next-generation space-based far-infrared telescopes.[8,9] This detector would also create possibilities for THz spectroscopic studies at the single-photon level, such as measurements of the THz emission from individual nanostructures.[10]

The detector we have studied consists of a superconducting titanium (Ti) nanobridge approximately 4 μm long, 350 nm wide, and 70 nm thick, with $T_c \approx 0.30$ K (Fig. 1). The Ti nanobridge spans contacts consisting of thick niobium (Nb) with $T_c \approx 8$ K. The fabrication process has been described previously.[4] The dimensions of the Ti nanobridge were chosen to have an impedance close to 50 Ω in the normal (non-superconducting) state to facilitate efficient high-frequency coupling.

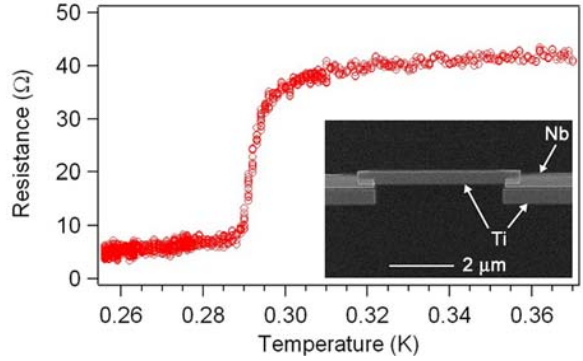

FIG 1. DC resistance as a function of temperature measured with 1 nA bias current. Inset: Scanning electron micrograph of Ti nanobolometer device on silicon substrate. The strips of Ti below the Nb contacts are an artifact of the fabrication process.

For photons with a frequency greater than the upper frequency scale for superconductivity in the Ti, $f_{Ti} \approx 3.5 k_B T_c/h$ = 22 GHz at $T \ll T_c$, the nanobridge impedance is approximately equal to the normal state resistance $R_n \approx 40$ Ω. In practice, the superconducting energy gap in the Ti is strongly suppressed by the bias current and temperature, so the relevant frequency scale is well below 22 GHz. The temperature rise due to an absorbed photon is $\Delta T = hf/C_e$, where f is the photon frequency, assuming that no energy is lost while the electron system reaches a thermal distribution. The larger superconducting energy gap in the Nb contacts, $\Delta_{Nb} \approx 1.2$ meV in our films, creates Andreev mirrors that inhibit the outdiffusion of heat from the Ti nanobridge.[3] The time for the initially excited photo-electron to share its energy with other electrons in the Ti and relax below $\Delta_{Nb}$ is $\tau_{e-e} \sim (2 \times 10^8 R_{sq} \Delta_{Nb} / k_B)^{-1} \sim 0.1$ ns, where $R_{sq}$ is the sheet resistance.[11] The initial excitations will spread a distance $\sim (D\tau_{e-e})^{1/2} \sim 0.1$ μm, where D is the diffusion constant, while the excitations cool to below $\Delta_{Nb}$. The subsequent energy removal is by electron-phonon coupling within the Ti, with an intrinsic thermal time constant $\tau_0 = C_e/G_{th} \sim \mu s$, where $G_{th}$ is the electron-phonon thermal conductance.[3,4]

A test system to study the detector response to single THz photons is under development but has presented significant technical challenges. A THz source coupled from outside the cryostat must be highly attenuated due to room temperature blackbody photons. Even with a source internal to the cryostat, the radiation power absorbed in the device must be ≲ fW to avoid exceeding the detector count rate. This requires carefully calibrated attenuation of the source and filtering of the out-of-band photon flux.

To facilitate rapid device characterization, we have developed an alternative testing technique that is easier to implement and avoids the problem of unwanted background photons (Fig. 2). The device is mounted in the light-tight inner vacuum can of a $^3$He cryostat with a base temperature of 230 mK. Absorption of a single THz photon is simulated by absorption of a 20 GHz microwave pulse with a duration of 200 ns, which is much shorter than $\tau_0$. We call this pulse a faux photon, or fauxton. The fauxton frequency $f_{fauxton} = E_{abs}/h$, where $E_{abs}$ is the absorbed energy of the microwave pulse, is adjusted simply by changing the amplitude of the microwave signal. The system coupling efficiency at 20 GHz is calibrated precisely above $T_c$ using Johnson noise thermometry, by comparing the temperature rise from a 20 GHz signal with the temperature rise from a known dissipated dc power. Since 20 GHz is greater than the frequency for superconductivity in the Ti, the impedance of the Ti nanobridge is approximately equal to $R_n$, as it is for an actual THz photon.

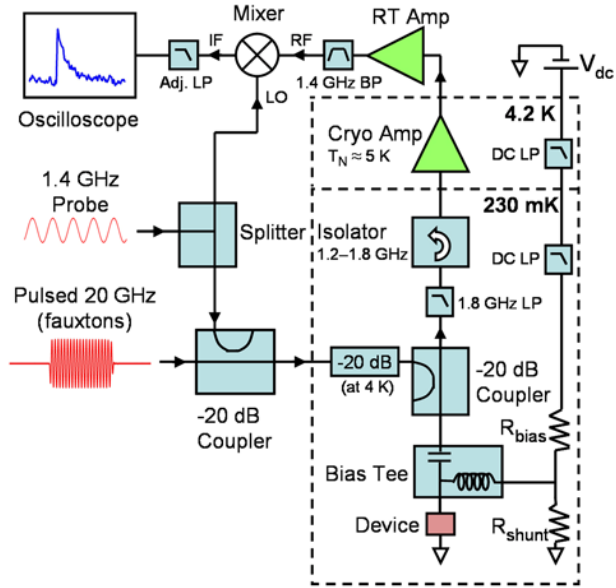

FIG 2. Schematic of experimental setup for fauxton testing. LP stands for low-pass filter and BP stands for band-pass filter. Some attenuators and filters have been omitted for clarity.

The resistance change when a fauxton is detected is recorded by measuring the change in the reflected power at 1.4 GHz. The 1.4 GHz probe signal reflected by the device is amplified using a low noise cryogenic amplifier ($T_N \approx 5$ K). The probe signal is amplified further and narrow band-pass filtered at room temperature, and then mixed with a phase-matched 1.4 GHz reference signal. The mixer output is low-pass filtered with an optimum bandwidth determined by the frequency crossover between thermal fluctuation noise and amplifier noise.[5] In our measurements, we chose 100 kHz as the bandwidth that empirically gave the best signal-to-noise ratio. This microwave measurement of the device impedance change takes advantage of the low noise cryogenic amplifiers and isolators available at these frequencies and avoids problems with electromagnetic pickup at lower frequencies.

The biasing condition is set by resistors mounted at the base temperature ($R_{bias}$ and $R_{shunt}$ in Fig. 2). The biasing line connects to the device through the dc port of a bias-tee, which has a bandwidth from dc - 5 MHz. We used $R_{bias}$ = 1 MΩ and $R_{shunt}$ = 50 Ω or 3 Ω. $R_{shunt}$ determines both the dc biasing condition as well as the load line seen at all frequencies relevant to the thermal response. The optimum dc bias point is like that of other superconducting bolometers.[12]

$R_{shunt}$ = 50 Ω corresponds approximately to the case of matched source and load impedances, in which case there is no electrothermal feedback.[13] In this case we should measure the intrinsic time constant, $\tau_0 = C_e/G_{th}$. We find $\tau_0$ = 7 μs, in good agreement with Ref. 4. With $R_{shunt}$ = 3 Ω, we have strong negative electrothermal feedback, which speeds up the device response.[6] The response time with strong negative electrothermal feedback, $\tau_{eff}$, depends on the bias point, $\tau_{eff} = \tau_0/(1 + L[R-R_{shunt}]/[R+R_{shunt}])$ with $R = V/I$ and $L = (dV/dI-R)/(dV/dI+R)$.[13] We find that the bias point with the optimum signal-to-noise ratio corresponds to a time constant of approximately 3.9 μs.

We next consider the detector energy resolution. At different fauxton frequencies, we measure a sequence of $10^3$ pulses with $R_{shunt}$ = 3 Ω and record each single-shot waveform. As an example, in Fig. 3 we plot a single-shot waveform and an averaged waveform for $f_{fauxton}$ = 50 THz. In the linear response regime and with no noise, the peak height is proportional to $f_{fauxton}$. We determine the peak height by averaging over a 2 μs window. We then make a histogram of the peak heights of all $10^3$ single-shot measurements for each fauxton frequency. The histograms are fit to a Gaussian function to extract the average peak height and the full-width at half-maximum (FWHM).

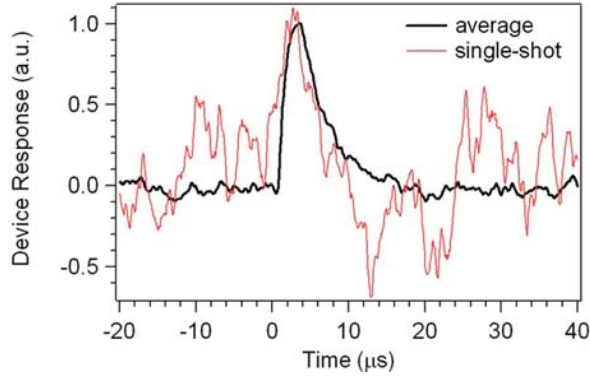

FIG 3. Average and single-shot device response to 50 THz fauxton.

In Fig. 4, we plot the histograms for fauxton frequencies of 25 and 50 THz, as well as for no fauxtons with the same 1.4 GHz probe power and the same bias point. We find that the best signal-to-noise ratio is obtained using a probe power that significantly reduces the critical current. The device response is linear with fauxton frequency, with a total full-width at half-maximum (FWHM) energy resolution $\delta E_{total}/h = 49 \pm 1$ THz. We also plot the histogram for no fauxtons with the bias current well above the superconducting critical current $I_c$, where the device is, to a good approximation, a temperature-independent resistor. In this case, the energy resolution should be limited by amplifier noise (plus a much smaller contribution from Johnson noise). We find $\delta E_{amp}/h = 43$ THz FWHM. The intrinsic detector noise and amplifier noise are assumed to be uncorrelated, hence $\delta E_{total}^2 = \delta E_{amp}^2 + \delta E_{intrinsic}^2$, giving $\delta E_{intrinsic}/h \approx 23$ THz FWHM.

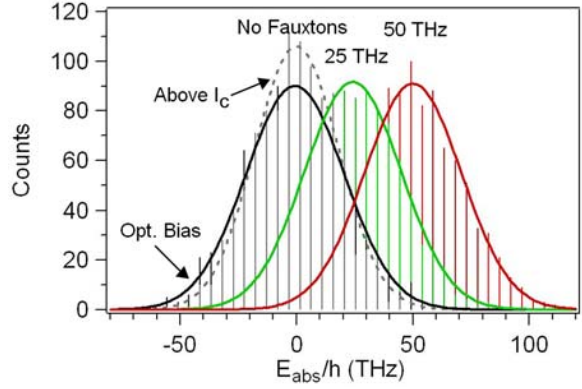

FIG 4. Histograms of single-shot device response to 50 THz fauxtons, 25 THz fauxtons, and no fauxtons. Response with no fauxtons is measured with the device optimally biased for detection. With the device above $I_c$, we measure the noise contribution from the amplifier.

The theoretical FWHM energy resolution is related to the noise equivalent power (NEP),[5]

$$\delta E_{th} = 2\sqrt{2\ln 2}\left(\int_0^\infty \frac{4df}{NEP^2}\right)^{-1/2}. \qquad (2)$$

Within the device response bandwidth, the dominant source of device noise is thermodynamic fluctuations, with a corresponding $NEP_{th}^2 = 4k_B T^2 G_{th}$.[5] We can estimate the intrinsic energy resolution by using $NEP_{th}$ in equation 2 with an upper limit of integration equal to the measurement bandwidth of 100 kHz. Using $G_{th} = 2.6 \times 10^{-12}$ W/K based on Ref. 4 and T = 0.3 K, we obtain $\delta E_{th}/h$ = 20 THz. This is in reasonable agreement with our experimental determination of the intrinsic energy resolution.

We also measured the output noise spectrum to determine if it is consistent with the measured energy resolution. In Fig. 5 we plot the noise power measured at the mixer input, expressed as a noise temperature referred to the input of the first stage amplifier. The noise was measured with no fauxtons at the optimum bias current and probe power, as well as with the bias current well above $I_c$. Well above $I_c$, the device is in the fully non-superconducting state and the noise is dominated by amplifier noise. At the optimum bias point, we fit the data to $T_N(f) = T_{amp} + T_0/(1+[2\pi(f-1.4\text{ GHz})\tau_{eff}]^2)$ with $T_{amp}$ the noise measured above $I_c$ and $T_0$ and $\tau_{eff}$ determined from the fit as 5.0 K and 3.9 μs, respectively.

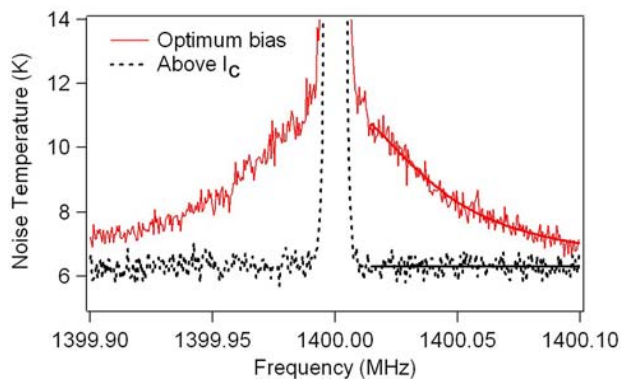

FIG 5. Measured noise spectrum at the mixer input, expressed as a noise temperature referred to the input of the first stage amplifier. The noise is measured both with the device at the optimum bias point and with the device in the non-superconducting state.

The device responsivity $S$ was determined by measuring the response to a square-wave-modulated 20 GHz excitation, with $S = 1.7 \times 10^7$ V/W. From the measured noise temperature and responsivity, we determine NEP(f).[14] Using this in equation 2, we predict $\delta E_{tot}/h$ = 50 THz with an upper integration limit equal to the measurement bandwidth of 100 kHz. This result is in good agreement with the measured energy resolution. We note that the frequency-dependence of the device noise is well described by the Lortentzian functional form expected for statistical thermal fluctuations, and it does not exhibit the excess noise seen in larger-area superconducting transition edge sensors.[15]

If the amplifier noise temperature were reduced to $T_N \lesssim 1$ K, as reported in recent studies,[16,17] the total energy resolution would be dominated by intrinsic device noise rather than by amplifier noise. The amplifier noise and the thermal fluctuation noise contributions to the energy resolution should both scale as the square root of the active device volume.[14] Hence future smaller devices should achieve an improved energy resolution by reducing the Ti nanobridge volume, and this improved energy resolution would not be limited by amplifier noise if the amplifier were satisfactory for the larger volume device.

Ultimately, the goal is the compare the fauxton technique to the detection of real THz photons. The fauxton technique avoids several significant complications of real photon detection, including imperfect optical coupling and the loss of energy from the initial photoexcitation due to outdiffusion or the emission of a high-energy phonon. Thus, the fauxton technique is not only a useful tool for preliminary device characterization, it can also be used for understanding detector non-idealities in real optical experiments.

The work at Yale was supported in part by NSF-DMR-0907082, NSF-CHE-0616875 and Yale University. The work by B.K. and S.P. was carried out at the Jet Propulsion Laboratory, California Institute of Technology, under a contract with the National Aeronautics and Space Administration. The work at Rutgers was supported in part by NSF-ECS-0608842 and the Rutgers Academic Excellence Fund. B.R. acknowledges Yale support from the Flint Fund for Nanoscience for research visits during the summers of 2008 and 2009. L.F. acknowledges partial support from CNR-Istituto di Cibernetica, Pozzuoli, Italy.